 \documentstyle[eqsecnum,aps,pre,epsf,tighten,subfigure]{revtex}

\def\ni{\noindent}
\def\th{\thinspace}
\def\eg{{\it e.g. }}
\def\ie{{\it i.e. }}
\def\cf{{\it c.f. }}
\def\viz{{\it viz }}

\def\bY{{\bf Y}}
\def\bX{{\bf X}}
\def\bx{{\bf x}}
\def\bF{{\bf F}}
\def\bg{{\bf g}}
\def\bG{{\bf G}}
\def\bB{{\bf B}}

\def\de{$d_E$}

\begin{document}

\title{Search for Low-Dimensional Chaos in Observational Data}

\author{J. R. Buchler}

\address
 {Physics Department, University of Florida, Gainesville,Florida, USA}

 \author{\vskip 5pt
 {\bf International School of Physics "Enrico Fermi", Course CXXXIII}
 \vskip -1pt
 {\bf "Past and Present Variability of the Solar-Terrestrial System:}
   \vskip -1pt
 {\bf Measurement, Data Analysis and Theoretical Models"}\\
 \vskip -1pt
 {Eds. G. Cini Castagnoli \& A. Provenzale.}
 } 

\maketitle

\section{Introduction}

Time-dependent signals reveal much more about their sources than do static
ones.  Thus, the period of a pendulum gives us information about the ratio of
the inertia and the restoring force.  Frequently, a Fourier spectral analysis
of a time-dependent signal displays steady peaks that are indicative of several
independent physical frequencies.  Such information is of course extremely
useful, not only in helping us model the phenomenon, but also in narrowing down
the model parameters.  The question naturally arises whether we might also be
able to extract similar constraints from {\sl unsteady temporal signals}.

All real life systems are very high dimensional (they have a very large number
of 'degrees of freedom'), such as well developed turbulence Often though, the
observed behavior is dominated by a few degrees of freedom.  In that case we
may represent these most important of these with a low dimensional dynamics and
hide our ignorance about the other ones by treating them as a stochastic
component.  In some cases we even get lucky and be able to ignore the
stochastic component all together.  This lecture concerns itself with this type
of situation.

Of course, in real systems one is further plagued by extrinsic physical and
observational noise.  How large a level of noise is admissible depends on many
factors, not the least of which is the nature of the dynamical system itself.

If the dynamics is stationary (autonomous system) and the signal is not
multi-periodic, then it is must be chaotic.  Sometimes the time-dependence is
much slower than the variability of the source, and for practical purposes we
may consider the system to be stationary.  We refer the student to the many
excellent introductions to chaos \cite{Berge},\cite{Ott},\cite{Thompson} that
abound in the literature.

Another difficulty of principle arises.  In variable star Astronomy, for
example, we generally have access only to the magnitude of the star (The
magnitude is defined as the log of the luminosity $m=-2.5 {\rm Log} L$).  Can
we ever hope to extract information about the D-dim dynamics from such a {\sl
scalar} time-series.  The answer is perhaps astonishingly, but fortunately,
affirmative.

Although the topic is relatively new there are good introductory books and
articles available on the problem of extracting a dynamics from a
time-series, for example \cite{Weigend} \cite{Abarbanel} \cite{Buchler_nyas96}.

\section{The Underlying Dynamics}

Suppose a physical system can be described by an autonomous system of ordinary
differential equations (ODEs) with
$d$ physical variables that make up the vector $\bY$.
 \begin{equation}
   {d{\bY}/dt} = {\bG} ({\bY}),        \label{Dyn}
 \end{equation}
 (A more complicated higher order system of ODEs can always be cast into this
canonical form).  We call this system a $d$-dimensional dynamics.  With given
initial conditions, the solution ${\bY}$ of these ODEs, i.e. the
$d$--dimensional vector of {\sl the physical phase space} that is referred to
as the trajectory in physical phase-space.

In most real situations we are unable to measure $\bY (t)$.  In fact, we
usually measure {\sl only one} quantity, say $m(t)$.  (In the case of variable
stars this would naturally be the luminosity or perhaps the radial velocity.)
This observable is generally a complicated and unknown function of $m(\bY )$ of
the physical state vector.

We assume that the discrete measurement times $\{t_n\}$ are equally spaced.
When this is not the case we have to interpolate which unfortunately introduces
additional noise.

The connection between $\bY (t)$ and $\{m_n=m(t_n)\}$ will be made through what
is termed {\sl a phase-space reconstruction}.  From $\{m_n\}$ we construct the
\de -uplets of vectors
    \begin{equation} {\bX (t_n) = \bX}^n 
      =\{m(t_n), m(t_n-\Delta), m(t_n-2\Delta), \ldots m(t_n-(d_E-1)\Delta) \}
    \end{equation}
 \ni where the delay $\Delta$ is an integer times the time-interval.  Both the
physical space and the reconstruction space are Euclidian ($d$ and \de\ are
integer).

 We assume now that these vectors themselves are generated by a \de -dim
dynamics
     \begin{equation}
       {d{\bx}/dt} = {\bg} ({\bx}),                     \label{gx}
     \end{equation}
 or equivalently by an \de-dim map
     \begin{equation}
       {\bX}^{n+1} = {\bF}[{\bX}^n],                     \label{FX}
     \end{equation}
 \ni It does not matter whether we use the description \ref{gx} or \ref{FX} --
the two are clearly related with $\{\bX^n=\bx(t_n)\}$.  It is natural to assume
that the observable $m$ is a smooth function of the physical phase space
variable $\bY$, \viz $\{m(t_n)\}=\{m({\bY}(t_n))\}$.  Thus, for example,
we assume that the luminosity of a star is a smooth function of the dynamical
variables, such as radii, velocities and temperatures, for example.

A theorem now states that there is a one-to-one (differentiable) relation
between the vector function $\bY (t)$ in the $d$-dim physical phase-space and
the vector function $\bx (t)$ n the \de-dim reconstruction space, provided that
\de\ is sufficiently large so that the reconstructed trajectory is devoid of
intersections and cusps.  When the latter condition is satisfied one says that
the reconstruction space is {\sl an embedding space}, and no additional benefit
can be gained from further increasing the dimension \de\ (A circle can
generically be embedded in 3-D; putting it in 4-D or 5-D does not change the
circle).  The theorem furthermore states that {\sl \de\ need not be larger than
2 times the fractal dimension of the attractor}, rounded off to the next
integer value \cite{Sauer}.  In practical time-series, that one need not go to
this bound and often $d_E=d$.  In the following we will loosely refer to the
dimension of the reconstruction space as \de.  Another obvious, but important
constraint is that the function $m(\bY)$ preserves all the necessary
information about all the degrees of freedom of the dynamics.

This theorem is extremely powerful because some properties are preserved in the
embedding, such as the Lyapunov exponents and the fractal dimension
\cite{Sauer} \cite{Abarbanel}.  Once the map has been obtained we can thus {\it
infer quantifiable properties of an unknown underlying dynamics from the
observations of a single variable}.  This is important because often an
observational data set is too short or too noisy to allow a {\it direct}
estimation of these quantities.  Of the fractal dimensions the Lyapunov
dimension $d_L$ is easiest to calculate, and it is defined in terms of the
Lyapunov exponents $\{\lambda_i\}$ as
    \begin{equation}
       d_{_{L}} = K + {1 \over \vert\lambda_{K+1}| } \sum_{i=1}^K \lambda_i
 \label{Lyap}
     \end{equation}
 \ni where the $\{\lambda_i\}$ are in ordered in decreasing size and $K$ is the
largest integer for which the sum is positive.

The {\it minimum} embedding dimension of course sets an upper bound on the
physical dimension $d$.  A lower limit is obtained from the obvious fact that a
fractal attractor has to be embedded in an integer dimensional space.  Thus
     \begin{equation}
          d_L< K+1 \leq d_{_{L}} \leq d_E          \label{Ineq}
     \end{equation}
 \ni When we are lucky the two limits coincide and we can obtain $d$.

\vskip 20pt

It is our aim to find a global representation, as opposed to local
representations, of the flow, i.e. a single function $\bF$ that evolves the
trajectory according to eq.~\ref{FX} in {\sl all} of phase-space.

We have no a priori knowledge about how to choose the global map ${\bf F}$ (or
the flow $\bg$).  It is however natural to try a polynomial expansion.  This is
particularly appropriate when the dynamics is only weakly nonlinear, as seems
to be the case for example for stellar pulsations \cite{Buchler_nato}.  Thus we
set

 \begin{equation}
   {\bf F}({\bf X}) = \sum_{k=1}^p {\bf C_k} P_k({\bf X}),       \label{Pmap}
 \end{equation}

\ni where the polynomials $P_j({\bf X})$ are constructed to be orthogonal on
the data set \cite{Brown}, \cite{Giona}, \cite{Buchler_prl} and $p$ is the
expansion order (typically 4 or 5).  

An equivalent approach is to treat the polynomial expansion as a least squares
problem, \ie one minimizes the sum ${\cal S}$
 \begin{eqnarray}
    {\cal S} &=& \sum_{\alpha=1}^{d_e} E_\alpha \\
    E_\alpha &=& \sum_n \bigl( X^{n+1}_\alpha -
F_\alpha(\bX^n)\bigr)^2\\  
    \bF(\bX) &\equiv& \sum_k \bB^k{\bf{\cal M}}_k(\bX)
 \label{error}
 \end{eqnarray} 
 \ni with respect to the vector coefficients $\bB^k$, where ${\bf{\cal
M}_k}(\bX)$ represents all the monomials up to order $p$ that can be formed
with the $d_e$ components $X_\alpha$ of the vectors $\bX$.  The sum over $n$
runs over the whole data set.
In the following we will also use one of the components $E_\alpha$ do measure
the error.

Instead of a standard QR least squares algorithm it is better to use a singular
value decomposition (SVD) approach to effect the minimization and to obtain the
coefficients of the polynomial map \cite{Serre_s2a}.  The SVD algorithm has
several advantages \cite{Num_rec}.  First, it is a very stable numerical method
and it is more than ten times faster than the Gram-Schmidt procedure.  Second,
it allows one to introduce a large number of unknown parameters in the fit
(even more than there are data points).  This is important because the number
of parameters goes up very fast with polynomial order $p$, as (${\rm
C}_p^{p+d_e}$) and other methods such as QR become singular.  With SVD the
redundant combinations give rise to eigenvalues that fall below the machine
error or a self-imposed cut-off $\omega_c$ value and they are {\it
automatically eliminated}.

\section {A Test case: the R\"ossler Oscillator}

The system of 3 equations known as the R\"ossler oscillator \cite{Thompson} is
given by

 \begin{eqnarray}
 \quad {dx_1\over dt} &=& -x_2 -x_3 \\
 \quad {dx_2\over dt} &=& x_1 +ax_2 \\
 \quad {dx_3\over dt} &=& b +(x_1 -c)x_3
 \label{roessler}
 \end{eqnarray}

\ni The dimension of the 'physical' phase-space of the R\"ossler system is thus
$d$=3.  The R\"ossler oscillator has been widely studied \cite{Thompson}
\cite{Packard}.  In some regions of parameter space ($a$, $b$, $c$) the
trajectory fills what is commonly a R\"ossler band (It is not really a band but
is made out of infinitely fine layers, somewhat like 'p\^ate feuillet\'ee').

The trajectory of this system is chaotic (\ie one of the Lyapunov exponents is
positive and nearby trajectories are divergent) and the attractor is
strange (it has a fractal dimension) \cite{Ott}.  Since strange and chaotic
behavior seem to occur simultaneously for the systems of interest to us we
shall simply refer to the attractor as chaotic.

\subsection{The test-data sets}

For our tests we have chosen the values $a$=0.2, \th $b$=0.2 \th and
$c$=4.8\quad in eq.~\ref{roessler} for which the solution is chaotic and the
attractor is in the shape of a band, the so called 'R\"ossler band'
\cite{Packard}\cite{Thompson}.  In the following, when we talk about the
R\"ossler band or attractor, we refer to the R\"ossler system with these
specific parameter values.  Other parameter values give attractors with
different properties.  For further details of all these tests we refer to Serre
et al. \cite{Serre_s2a}.

On top in Fig.~1 we show the temporal segment of $x_1(t)$ that we use as test
input data in the nonlinear analysis.  (The data sequence has been obtained by
integrating the R\"ossler system with a fourth order Runge-Kutta algorithm with
a fine time-step of 0.001.)  Since we want to test our method we have
specifically chosen a relatively short test-data set of 4000 points with
$\delta t$ chosen so as to give of the order of 60 equally spaced points per
'cycle'.

We stress that this segment of $x_1$ data which is sampled at regular
time-intervals, is all the information that we allow ourselves in the global
flow reconstruction, \ie {\it we do not use any information about the other two
variables}, $x_2(t)$ and $x_3(t)$.

It has been shown that some nonlinear methods can erroneously indicate low
dimensional behavior for stochastic data (\cite{Provenzale}\cite{Smith} also
cf. the lectures by Provenzale and by Smith in this Volume).  It is therefore
important that our analysis be able to clearly distinguish between the first,
deterministic, and the second, stochastic data sets.

We have thus also created a second, stochastic test-data set as follows
\cite{Serre_s2a}.  We first 'pre-whiten' the R\"ossler data set by subtracting
the dominant frequency and its 8 harmonics, all with their respective Fourier
amplitudes and phases -- we thus split the signal into to a periodic and an
irregular part.  The Fourier spectrum of the irregular component has a broad
band structure, and we compute the envelope of its (complex) Fourier spectrum.
Then we convolve its inverse Fourier transform with a white noise signal, in
other words we create a stochastic signal with the same color as the irregular
signal.  Finally we mix this colored stochastic signal with the originally
subtracted periodic component of the R\"ossler signal.  The resultant
stochastic data set is displayed in Fig.~1.  Note that, by construction, the
envelope of the Fourier spectrum of the stochastic signal is the same as that
of the R\"ossler signal.

 \begin{figure}
  \begin{center}
      \epsfysize=6.cm
      \leavevmode
      \epsfbox{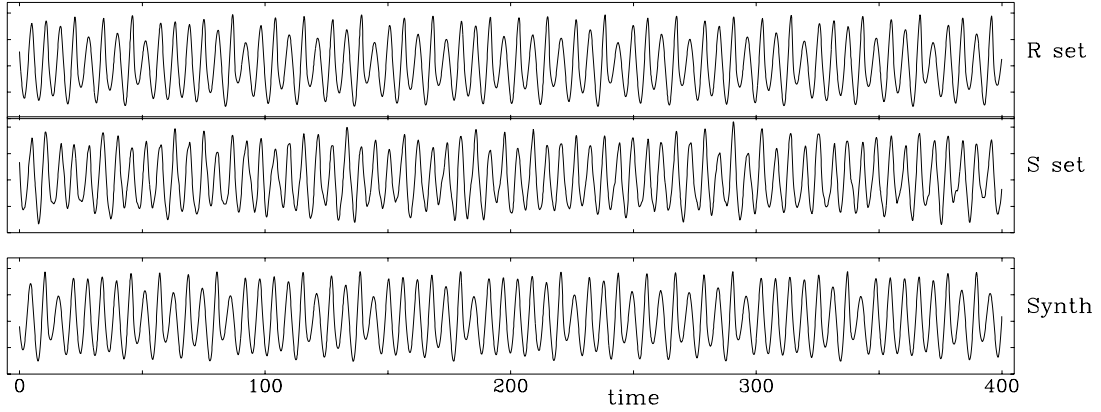}
      \caption{Input data sets: 
        {\it top}: R\"ossler data; {\it middle}: stochastic data, 
         Reconstructed signal: {\it bottom}: stochastic data}
  \end{center}
 \end{figure}

By eye it is impossible to distinguish between the R\"ossler and the stochastic
data sets.  The Fourier spectra do not distinguish either because, by
construction, they are very similar.  What tools do we have to distinguish
between the two signals?

The study of many systems has shown that it is often revealing to make a 'first
return map', i.e. to plot one extremum of the signal versus the following one
\cite{Thompson}\cite{Packard}.  Fig.~2 compares the first return map for the
R\"ossler (R) and the stochastic (S) signals.  Clearly the S signal does not
have the coherence of the R signal.  The close resemblance of the R map to the
logistic map \cite{Ott} suggests that the R\"ossler attractor is born out of a
cascade of period doublings and that a horseshoe structure underlies its
dynamics.  We caution though that only very dissipative attractors can give
rise to such nice compact first return maps.

 \begin{figure}
    \begin{center}
      \epsfysize=4.5cm
      \leavevmode
      \epsfbox{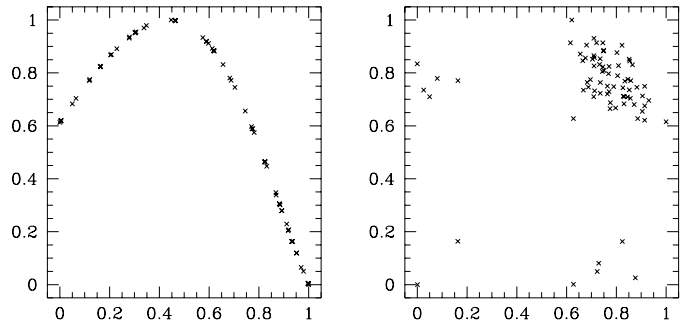}
      \caption{First return maps: 
            {\it left}: R\"ossler data; {\it right}: stochastic data}
    \end{center}
 \end{figure}

While it is of course possible to plot the trajectory in a 3-D picture, this is
generally not the optimal type of representation because often the attractor is
compressed in some directions.  It is preferable to project the trajectory onto
Broomhead and King \cite{Broomhead} coordinates $\{\xi_i\}$.  These projections
onto the eigenvectors of the correlation matrix have the property of providing
an optimal spreading of the attractor in all 'directions'.  The first column of
Fig.~3 shows the BK projections onto the first three axes for the R signal and
the fourth column that of the S signal.  Here we see again that the stochastic
signal is fuzzier (less coherent).

 \begin{figure}
    \begin{center}
      \epsfysize=6.5cm
      \leavevmode
      \epsfbox{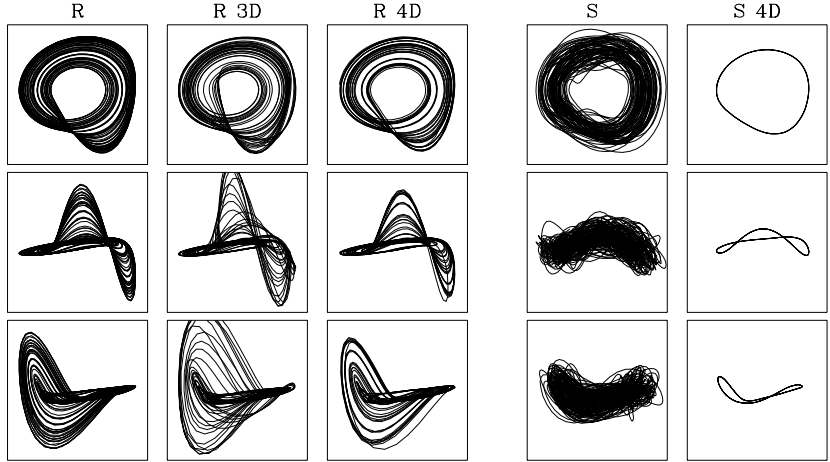}
      \vskip 5pt
      \caption{Broomhead-King projections; {\sl top}: $\xi_2$ vs $\xi_2$,
       {\sl middle}: $\xi_1$ vs $\xi_2$, {\sl bottom}: $\xi_2$ vs $\xi_3$;
        R\"ossler data, col.~1; Reconstructed signal col.~2 3-D, col.~3 4-D;
      Stochastic data, col.~4,  Reconstructed signal, col.~5}.
    \end{center}
 \end{figure}

Abarbanel et al. \cite{Abarbanel} have developed a simple, yet very useful and
powerful tool for estimating the embedding dimension, namely the method of
'false nearest neighbors' \cite{False}.  The underlying principle is best
described by the following analogy: If we defined the distances between stars
as being those which we measure on the celestial sphere we would find many
nearest neighbors that are 'false' because they merely fall on neighboring
lines of sight.  We have to go to higher dimension (3 here) to correctly
determine the distances.  The method of false nearest neighbors generalizes
this notion.  One computes the distances between points $\bX_n$ in higher and
higher embedding dimensions \de.  When the percentage of nearest neighbors does
not change any more with increasing \de\ we have found the minimum embedding
dimension.

Fig.~4 displays the percentage of false nearest neighbors as a function of \de\
and the tolerance (\cite{Abarbanel}), and for two values of the delay parameter
$\Delta$.  The percentages drop off very fast for the R set that is displayed
in the two figures on the left, and they indicate that the R data are
embeddable in 3-D.  The situation is quite different for the S data set shown
in cols.~3 and 4, indicating a much larger dimension (as it should because the
dimension of a stochastic signal is indeed very large; the percentage still
decreases with \de\ because for the small number of data points the density of
points decreases so fast with dimension \cite{Abarbanel}).

 \begin{figure}
    \begin{center}
      \epsfysize=4.5cm
      \leavevmode
      \epsfbox{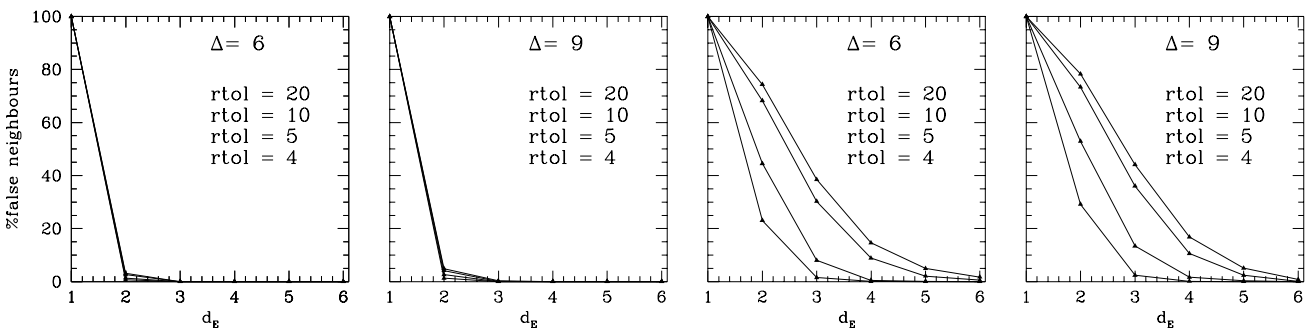}
      \caption{Percentage of false nearest neighbors: 
            {\sl left}: R signal, {\sl right}, S signal}
    \end{center}
 \end{figure}

A totally different characterization of an attractor, of a topological nature,
has been discussed in Letellier et al. \cite{Letellier} and, as an example, has
been applied to the R\"ossler band as well.  The idea is that a chaotic
attractor can be described by the population of unstable periodic orbits, their
related symbolic dynamics and their linking numbers.  In three dimensional
cases, periodic orbits may be viewed as knots and, consequently, they are
robust with respect to smooth parameter changes allowing the definition of
topological invariants under isotopy (continuous deformation).  This
topological approach is thus based on the organization of periodic orbits and
symbolic dynamics.  It is a very powerful tool, but unfortunately, it only
works for attractors for which the first return map consists of relatively well
defined branches.

\subsection{The Flow Reconstruction}

We are ready now to discuss the global polynomial flow reconstruction.  Our
reconstruction contains several unknown parameters: First we have the dimension
\de.  It is of course the minimal value of this parameter that we would like to
determine, and we increase \de\ until we obtain an embedding.  Next we have the
delay $\Delta$.  Studies of nonlinear time-series have shown that for many
purposes there is an optimal time-delay $\Delta$ that is obtained from mutual
information considerations (\eg \cite{Weigend} \cite{Abarbanel}).  However, for
our analysis this value is not necessarily the best.  For example, too short a
delay collapses the attractor onto the diagonal and noise becomes a serious
problem, whereas too large a delay causes a very strong nonlinearity in the
map, such that a polynomial assumption cannot handle it.  We therefore treat
$\Delta$ as a free parameter in our search and anticipate (hope) that there is
some range in which a polynomial map works fine.  Finally, the parameter $p$ is
the highest order polynomial that we allow in the map.  Again since the number
of coefficients of the map goes like C$^p_{d_e+p}$ we would like to keep it to
a minimum.

 \begin{figure}
    \begin{center}
      \epsfysize=5.5cm
      \leavevmode
      \epsfbox{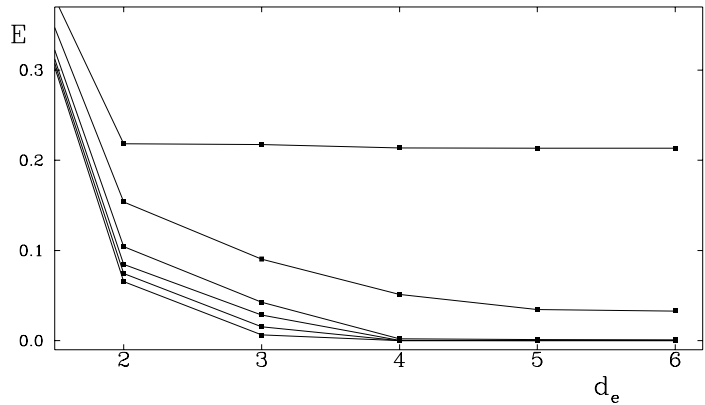}
      \caption{Error norm as a function of embedding dimension
                \de\ and polynomial order, $p$=2, $\ldots $ 6, increasing 
                 from top to bottom}
    \end{center}
 \end{figure}

In Fig.~5 we display a typical behavior of the error norm as a function of \de\
and of $p$ for fixed delay $\Delta$=11.  We only show the error for the first
component $\alpha$=1 of $\bX$ (\cf eq.~\ref{error}) because the other
components behave similarly.  As one would expect the error falls off
monotonically with polynomial order $p$.  In 3-D the error norm decreases
relatively slowly though.  A probable reason is that a low order polynomial
approximation to the R\"ossler map may not be accurate enough (A rational
polynomial map might be more appropriate, \cite{Brown} \cite{Gouesbet}).

The important result from Fig.~5 is that the norm shows a very clear drop and
levelling off at $d_e$ around 3 and 4.  This suggests that we should be able to
construct a global map in 4-D, but that it may already be possible to do so in
3-D.  This evidence alone is however not sufficient to establish the minimum
embedding dimension, and we need to look also at the behavior of the properties
of the {\sl synthetic trajectories}.  By this we mean trajectories $\{\bX^n\}$
that, once we have constructed the map $\bF$, we obtain by iteration of
eq.~\ref{FX} starting with some seed value $\bX^0$.  We refer to any {\it
component} $X^n_\alpha$ of these trajectories as a {\it synthetic signal}.
Indeed it might be that the error norm is very small, indicating a good fit,
but that the generic trajectories of the map are very different from the test
data.

Now a word about Lyaponov exponents (cf.~e.g. \cite{Ott}) which give a useful
quantitative measure of the map and of its attractor.  In the case of the
R\"ossler system where we know the 3 ODEs we can compute the 3 Lyapunov
exponents in our chosen ($a$, $b$, $c$) regime directly -- They are found to
have the values $\lambda_1$=0.057, $\lambda_2$=0, $\lambda_3$=--4.5, and
$d_{_{L}}$= 2.013.  One notes that one of the Lyapunov exponents of the
R\"ossler system is exactly zero because the system of ODEs is autonomous (a
time independent flow) \cite{Ott}.  Next, the presence of a {\sl positive}
Lyapunov exponent indicates that the attractor is {\sl chaotic}.  Finally, the
fractal Lyapunov dimension implies that it is a strange attractor \cite{Ott}.

The question now arises whether it is possible to recover the values of these
Lyapunov exponents from the R test data?  The answer is two-fold: First, the
computation of all of the Lyapunov exponents from a time series necessitates
{\it long} data sets, often at least an order of magnitude longer than is
available from observations.  Here, for example, the length of our R test-data
set (4\th 000) is far too small to yield exponents that come even close to the
correct values.  In observational data, in addition, noise easily vitiates the
reliability of the Lyapunov calculations.  Second, however, we can first
construct the map and then generate synthetic trajectories of sufficient length
to be able to compute the Lyapunov exponents.  The reason why, with this two
step procedure, we are not 'getting something for nothing' is discussed in
ref. \cite {Buchler_s3}.

Table~1 compares these Lyapunov exponents and Lyapunov dimension for synthetic
trajectories for various values of embedding dimension and time-delay.  In
$d_e$$\geq$4 the largest Lyapunov exponents ($\lambda_1$) are seen to be very
close to the exact value.  An exception occurs for $d_e$=3 where $\lambda_1$
and $\lambda_3$ are too large by a factor of about 2.  This fact is probably
connected with the relative lack of stability of the map in 3-D.  In the next
section we shall see that the addition of small amounts of noise stabilizes the
map and produces correct Lyapunov exponents.  The scatter of the negative
exponents is higher than for the positive ones, but this is to be expected
\cite{Sauer}.  Table~1 shows that, except for \de = 3, the Lyapunov dimensions
are close to their known exact values.  For the sampling rate that we use here
the exponents are essentially the same whether we compute them for the flow or
for the map.

We draw attention to the fact that $d_{_{L}}$ {\it is independent of the
embedding-dimension $d_e$}.  If our attractor reconstruction is to make sense
this is of course a necessary condition, and conversely it is a check on our
calculation.

The second Lyapunov exponent $\lambda_2$ is always very small ($\approx
10^{-4}$).  Although we have constructed a map here, the sampling-time is
sufficiently small so that the map closely approximates a flow.  The smallness
of one Lyapunov exponent constitutes a very powerful check on our {\sl a
priori} assumption that a flow(eq.~\ref{FX} indeed underlies the dynamics.

As expected there is a range of delays for which one obtains good maps.  In
Fig.~1, bottom, we show the best synthetic signal (obtained with \de\ = 4, $p$
= 4).  Is has very similar properties to the original set.  (Because the
time-series are chaotic, it is not very meaningful to compare them directly.
Since the signals are so short, statistical comparison methods are not very
useful here either.  We therefore have to rely on a number of comparison tests,
neither of which is totally convincing by itself, but when taken together they
give a good level of confidence \cite{Serre_s2a} \cite{Buchler_s3}.  We note
here that the Fourier spectrum of the synthetic signal is very similar to that
of the R data set.  The Broomhead King projections (Fig.~3) are also very
similar.

In contrast, the flow reconstruction of the S test data set yields only limit
cycles, for all values of $\Delta$ and $p$.  In fact the limit cycles are
almost identical to the periodic part of the signal.  The global flow
reconstruction thus recognizes the stochastic noise for what it is and it does
not erroneously indicate low dimensional behavior.

\subsection{Noise}

The reconstruction of a stable map encounters some difficulties for the
R\"ossler data set.  The reason behind it is that the neighborhood of the
attractor is not well described.  One can understand this as follows: Suppose
we tried to apply this method to system which had a limit cycle.  Unless we had
a transient we would have no information whatsoever on the neighborhood of the
limit cycle, and the neighborhood could equally well come out repulsive or
attractive.  Noise broadens the attractor and forces the map to form a stable
neighborhood.

 \vskip 15pt
 \begin{center}
  \begin{tabular}{rrrrrrrrr|rrrrrrrrr}
  \hline\hline\noalign{\smallskip\smallskip}
  \multicolumn{18}{c}{TABLE 1.\quad Lyapunov exponents and dimension, 
    {\sl left:} R set, 
    {\sl right:} R set + 0.2\% gaussian noise}~~~~~~~~~~~~\\
  \noalign{\smallskip\smallskip}
  \hline\hline\noalign{\smallskip\smallskip\smallskip}
 \de~~ & ~~$\Delta$~~& ~~$\lambda_1$~~ & ~~$\lambda_2$~~
    &  ~~$\lambda_3$~~ & ~~$\lambda_4$~~ & ~~$\lambda_5$~~
    & ~~$d_L$~~ &  &  &
 \de~~ & $\Delta$~~ & ~~$\lambda_1$~~ & ~~$\lambda_2$~~\
    & ~~$\lambda_3$~~ & ~~$\lambda_4$~~ & ~~$d_L$~~\\
 \noalign{\smallskip}
 \hline
 \noalign{\smallskip\smallskip}
 \multicolumn{18}{l}{exact values:}\\
 \noalign{\smallskip}
  \hline\noalign{\smallskip}
  &     & 0.057 &    0~~    & --4.5~  &  &        & 2.013& & &
  &    &  &  &           &        &   \\
  \hline\noalign{\smallskip\smallskip}
 \multicolumn{18}{l}{computed from synthetic signals:}\\
 \noalign{\smallskip}
  \hline\noalign{\smallskip}
3~~ & 5~~  & 0.128\quad & $\quad <$$10^{-4}$ & --6.4~  & &  & 2.020 & & &
  3~~ & 4~~ & 0.046 & $\quad <$$10^{-4}$ & --17.8~  &        & \ 2.003  \\
4~~ & 4~~ & 0.057 &           & --3.9~  & --16.7~ &        & 2.014
                    &\quad\quad&\quad\quad &
  3~~ & 5~~ & 0.041 & & --16.9~ &        & \ 2.003  \\
4~~ & 5~~ & 0.058 &           & --3.7~  & --15.0~ &        & 2.014& & &
  3~~ & 6~~ & 0.055 & & --16.2~ &       &  \ 2.003  \\
4~~ & 6~~ & 0.051 &           & --3.5~  & --12.7~ &        & 2.014& & &
  3~~ & 7~~ & 0.062 & & --14.3~ &       &  \ 2.004  \\
4~~ & 7~~ & 0.068 &           & --3.2~  & --9.0~  &        & 2.021& & &
  4~~ & 4~~ & 0.063 & & --17.3~ &        --30.5 &  \ 2.004  \\
5~~ & 4~~ & 0.054 &           & --8.0~  & --19.0~ & --28.0~ & 2.007& & &
  4~~ & 5~~ & 0.061 & & --16.8~ &       --28.4 &  \ 2.004  \\
5~~ & 5~~ & 0.054 &           & --6.8~  & --18.2~ & --28.0~ & 2.008& & &
  4~~ & 6~~ & 0.066 & & --16.8~ &        --28.5 &  \ 2.004  \\
5~~ & 6~~ & 0.055 &           & --5.3~  & --16.8~ & --29.0~ & 2.010& & &
  4~~ & 7~~ & 0.068 & & --15.4~ &         --26.4 &  \ 2.005  \\
5~~ & 7~~ & 0.055 &           & --4.8~  & --15.8~ & --29.0~ & 2.011& & &
   &      &       &        &           &        & \\ 
 \noalign{\smallskip}
 \hline\hline
 \end{tabular}
 \end{center}

\vskip 20pt

We conclude from this study (1) that the minimum embedding dimension is 3
and (2) that $K = {\rm int}(d_L)$.  From the inequality (\ref{Ineq}) it thus
follows that $d=3$.  In other words the global flow reconstruction method
recovers the dimension of the dynamical system (\ref{roessler}) that has
generated it.

\section {A Real system: The Light Curve of R Scuti}

In 1986 numerical hydrodynamic simulations of metal poor Cepheid models,
called W ~Virginis stars, showed irregular pulsations \cite{Buchler_chaos}
\cite{Kovacs_chaos}\cite{Buchler_nyas90}.  A further investigation revealed
period doublings as a stellar parameter was varied and concluded that the
irregular behavior of these models were a manifestation of low dimensional
chaos.  These calculations were based on models that ignored convective energy
transport and treated radiation transfer in an equilibrium diffusion
approximation.  In order to confirm the theoretical prediction as to the
chaotic nature of the irregular variability it was obviously necessary to
compare to observations.

Observationally, it has been known for a long time that the light-curves of W
Vir stars are regular (periodic) for low periods, but then develop
irregularities as the period and luminosity increase.  This increasingly
irregular behavior carries over to their more luminous siblings, the RV Tauri
stars.  Unfortunately, these irregular variables have not received much
attention from professional astronomers - the publication of what {\sl appears}
to be a table of random numbers is not compatible with career and funding
pressures!  Hopefully, the development of nonlinear time-series analyses such
as this one will change this perception.  However the Association of Variable
Star Observers (AAVSO) in Cambridge collects observations and compiles data
from amateur astronomers around the globe.  Often these data are visual, and of
poor quality.  Yet, the large number of independent measurements make this a
wonderful data base.  For the variable star R Scuti a large data set of 15
years of data are available.  The data have a normal distribution that is
independent of magnitude, and it is therefore possible to extract a good
light-curve despite the large individual errors.  We show a short section of
the observational data of R Sct in Fig.~6 together with a smooth fit.  It is
this fit with equally spaced data points that we use as the input data for our
global flow reconstruction.  For more details of the smoothing procedure we
refer the reader to the original papers \cite{Buchler_prl} \cite{Buchler_s3}.
The whole R Sct light-curve is displayed in Fig.~7 on top.

  \begin{figure}
    \begin{center}
          \epsfysize=4.cm
      \leavevmode
      \epsfbox{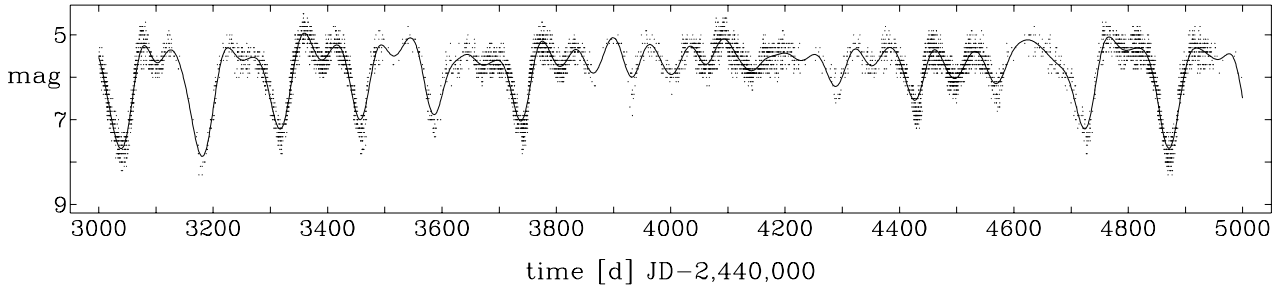}
      \caption{Typical section of the observational data together with a smooth
                 fit}
    \end{center}
 \end{figure}

The Fourier spectral analysis of the light-curve shows a broad dominant peak at
$f\approx 0.007$ d$^{-1}$ and a secondary much broader structure centered on
$f\approx 0.014$ d$^{-1}$.  The analysis of patchy, but very old data going
back some 150 years has shown that while the envelope of the Fourier spectrum
is more or less steady, the individual peaks are not at all, especially the
tertiary ones \cite{Kollath}.  There is a further discussion in
ref.~\cite{Buchler_s3} as to why the light-curve of R~Scuti cannot be generated
by multiperiodic pulsations, even when evolutionary changes are taken into
account.

Space does not permit us to describe the details of the global flow
reconstruction and we refer the reader to the original papers
\cite{Buchler_prl} \cite{Buchler_s3} (see also \cite{Buchler_nyas95}).

In Fig.~7, middle and bottom, we exhibit two pieces of synthetic signal
(obtained in \de\ = 4 with $p$ = 4).  Visually, the synthetic time-series is
very similar to the observational data.  The Fourier spectrum similarly has the
same envelope.

 \begin{figure}
    \begin{center}
      \epsfysize=6.5cm
      \leavevmode
      \epsfbox{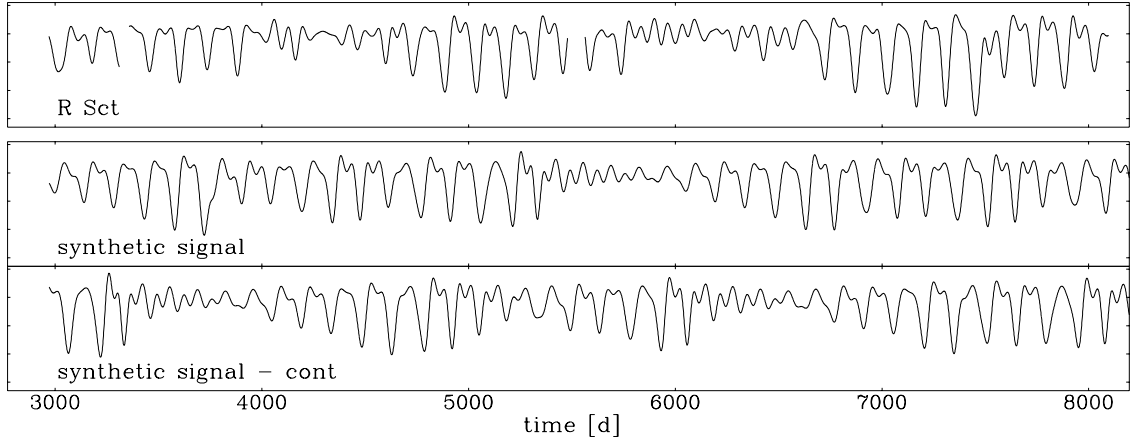}
      \caption{R Scuti, {\sl top}: Whole light-curve data set used in the 
     analysis, {\sl middle and bottom}: pieces of synthetic light-curves
generated with the global map}
    \end{center}
 \end{figure}

In Fig.~8 we show the Broomhead-King projections on the lowest three axes.  The
left-most column represents the observational data.  Columns 2 and 3 show our
best reconstruction with 4-D and 5-D maps (eq.~\ref{FX}).  Finally, in col.~4
we show a reconstruction with a flow (eq.~\ref{gx}).

In Table~2 we show the Lyapunov exponents and dimension for several values of
$\Delta$ and \de.  Noteworthy are the following points: (1) the results show a
certain robustness with respect to the delay $\Delta$, (2) $d_L\approx 3.1$ is
essentially independent of \de, (3) $\lambda_1 > 0$ which establishes the
chaotic nature of the attractor, and (4) $\lambda_2 \approx $ 0, which confirms
the presence of a flow.  We note that it has been found impossible to construct
a map with synthetic signals that bear any resemblance to the data which
indicates that the minimum embedding dimension must be greater than 3.
However, as Table~2 shows in 4-D we can construct a flow whose propeties remain
largely invariant as we increase \de.  We conclude that the minimum \de\ is
therefore 4.  We add in passing that a false nearest neighbor analysis
corroborates 4 as the minimum embedding dimension.

 \begin{figure}
    \begin{center}
      \epsfysize=6.5cm
      \leavevmode
      \epsfbox{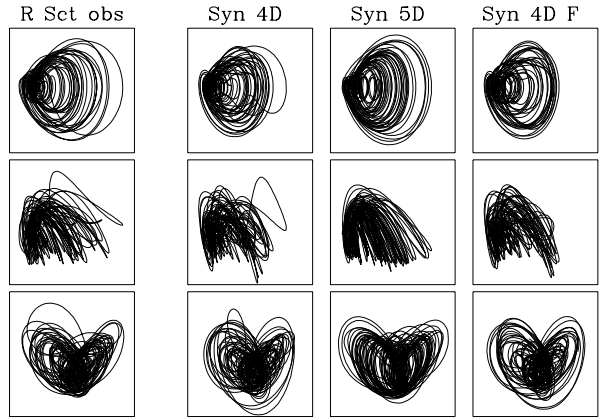}
      \caption{Broomhead-King projections; {\sl top}: $\xi_2$ vs $\xi_2$,
       {\sl middle}: $\xi_1$ vs $\xi_2$, {\sl bottom}: $\xi_2$ vs $\xi_3$; {\sl
left:} R Scuti observational data, {\sl cols. 2 and 3}: map reconstructions in
4-D and 5-D, {\sl right:} flow reconstruction in 4-D;
  }    \end{center}
 \end{figure}

Having now established from the Lyapunov dimension that the lower bound $K$+1
\th (rel.~\ref{Ineq}) of the physical phase-space dimension is 4 we thus
conclude that $d=4$, i.e. the light-curve of R Scuti is generated by a 4-D
dynamics.

 \vskip 10pt
 \begin{center}
 \begin{tabular}{cccccccc}
  \hline\hline
  \noalign{\smallskip\smallskip}
  \multicolumn{8}{c}{TABLE 2.\quad Lyapunov exponents and dimension, R Scuti}\\
  \noalign{\smallskip\smallskip\smallskip}
  \hline\hline
  \noalign{\smallskip\smallskip\smallskip}
  ~~\de~~ &  ~~$\Delta$~~ & ~~p~~ & $\lambda_1$ &  $\lambda_2$ 
     &  $\lambda_3$  &  $\lambda_4$  & $d_L$ \\
 \noalign{\smallskip}
 \hline
 \noalign{\smallskip}
  4&  4 & 4 & 0.0019 &~~~$<$$10^{-4}$ & ~~--0.0016~~ & ~~--0.0061~~ &~~3.05~~\\
  4&  5 & 4 & 0.0017 &            & --0.0014 & --0.0054 & 3.06 \\
  4&  6 & 4 & 0.0019 &            & --0.0009 & --0.0051 & 3.19 \\
  4&  7 & 4 & 0.0020 &            & --0.0011 & --0.0052 & 3.18 \\
  4&  8 & 4 & 0.0014 &            & --0.0010 & --0.0049 & 3.07 \\
  5&  7 & 3 & 0.0016 &            & --0.0005 & --0.0041 & 3.27 \\
  6&  8 & 3 & 0.0022 &            & --0.0003 & --0.0018 & 3.52 \\
 \noalign{\smallskip}
 \hline\hline
 \end{tabular}
 \end{center}

 \section{Conclusion}

The polynomial global reconstruction method has been tested on the R\"ossler
attractor.  Using as input a short time-series of only one of the 3 R\"ossler
variables one obtains a very good estimation of the Lyapunov exponents and the
fractal dimension of the attractor.  Furthermore it is possible to determine
that this scalar signal is generated by a 3-D flow (the R\"ossler equations
\ref{roessler}), in other words one can find out the dimension of the physical
phase-space of the dynamics.

Encouraged by these results the flow reconstruction has been applied to real
data, viz the light-curve of the variable star R~Scuti.  Not only are these
data contaminated with a good amount of observational and other extrinsic
noise, but we have no a priori information about the dynamics that has
generated them.

The analysis has shown that the dynamics which generates the light-curve of
R~Scuti is of dimension four, i.e. the signal can be generated by a flow in 4-D
(eq.~\ref{Dyn}).  Furthermore the attractor has a fractal dimension $\approx $
3.1.  This must be considered remarkable considering that this star undergoes
luminosity fluctuations of up to factors of 40 (cf Fig.~1), with shock waves
and ionization fronts criss-crossing the stellar envelope.  Some of the
physical implications have been addressed in references \cite{Buchler_prl}
\cite{Buchler_s3}.

If nothing else it is interesting that the nature of the irregular variability
of these types of stars is no longer a mystery, but that it has such a simple
explanation in terms (very) low dimensional chaos.

 \acknowledgments

It is a pleasure to acknowledge the collaboration of Thierry Serre and of
Zoltan Koll\'ath in the development and application of the global flow
reconstruction method.  This work has been supported by the National Science
Foundation.

 \end{document}